\documentclass[11pt]{article}
\usepackage{amssymb}
\usepackage{epsfig}
\begin{document}
\title{{\bf{\Large Entropy function from the gravitational surface action for an extremal near horizon black hole}}}
\author{ 
 {\bf {\normalsize Bibhas Ranjan Majhi}$
$\thanks{E-mail: bibhas.majhi@iitg.ernet.in}}\\ 
{\normalsize Department of Physics, Indian Institute of Technology Guwahati,}
\\{\normalsize Guwahati 781039, Assam, India}
\\[0.3cm]
}
\maketitle

\begin{abstract}
  It is often argued that {\it all the information of a gravitational theory is encoded in the surface term of the action}; which means one can find several physical quantities just from the surface term without incorporating the bulk part of the action. This has been observed in various instances; e.g. derivation of the Einstein's equations, surface term calculated on the horizon leads to entropy, etc. Here I investigate the role of it in the context of entropy function and entropy of extremal near horizon black holes. Considering only the Gibbons-Hawking-York (GHY) surface term to define an entropy function for the extremal near horizon black hole solution, it is observed that the extremization of such function leads to the exact value of the horizon entropy. This analysis again supports the previous claim that there exists a ``{\it holographic}'' nature in the gravitational action -- surface term contains the information of the bulk.   
\end{abstract}

\section{\label{Intro}Introduction and Motivation}
    The computations of entropy of an extremal near horizon black hole (ENH-BH) solution, proposed by A.$~$Sen, is a very simple and useful method \cite{Sen:2005wa} (For a review and large amount of references in this direction, see \cite{Sen:2007qy}). In this method, a function is introduced by integrating the Lagrangian density of the theory on the horizon of the ENH-BH geometry. Then performing the Legendre transformation of this function with respect to electric field strengths an entropy function is defined. Finally, the entropy is given by the extremum value of the entropy function with respect to the fields, appearing in the theory. One of the important facts of this approach is that the calculation of the entropy ultimately boils down to the solutions of a set of simplified algebraic equations.
One must note that the whole analysis was based on the action of the theory in which the gravitational part does not contain the surface term, like the Gibbons-Hawking-York (GHY) in case of the general theory of relativity (GR) -- both in terms of the definition of the initial function which is the integrated version of the Lagrangian density and in the expression for the horizon entropy (done by the Wald's Noether charge prescription \cite{Wald:1993nt,Iyer:1995kg} corresponding to the gravity action without the boundary term).

   In this paper, I shall develop the entropy function formalism alone from the surface term of the gravitational action in presence of the matter action of the theory. The computation will be confined within GR and hence the convenient boundary term will be taken to be the GHY term. Here I will not use any ``direct'' information of the Einstein-Hilbert (EH) part. GHY comes into the picture to obtain a well prescribed action principle in the derivation of the Einstein's equations of motion. Of course this choice is not unique \cite{Charap:1982kn}, but it becomes popular because its simplicity and wide applicability. The interests purely on the surface part are due to the following reasons. (i) As in the local region of spacetime Christoffel symbols vanishes, the EH action reduces to a purely surface term. (ii) Extremization of the surface term for a diffeomorphism, in which the diffeomorphism vector satisfies the constant norm condition, leads to Einstein's equations of motion \cite{Padmanabhan:2004fq}. (iii) Evaluation of the surface term on the horizon yields the entropy. This is done by calculating first on a constant radial coordinate surface and then taking the horizon limit \cite{Paddy}. Moreover the Noether charge corresponding to the GHY term, calculated on the horizon (similar to the Wald charge associated with the EH action), leads to the entropy \cite{Majhi:2012tf,Padmanabhan:2012bs}. This has also been even tested successfully in the context of Virasoro algebra and Cardy formula \cite{Majhi:2012tf,Majhi:2012nq}. (iv) In literature, often argued that the entropy is associated to the degrees of freedom around or on the relevant null surface rather than the bulk geometry of spacetime. All these instances indicates that either the surface term encodes all the information about the bulk or the surface action bears the dynamics of the system.

  Considering the above facts, it is important to investigate if entropy function formalism can be developed purely from the surface or the boundary term of the action. In this paper I shall show that it is possible. The steps are identical to the original work \cite{Sen:2005wa,Sen:2007qy}. The organization of the paper is as follows. In section \ref{GHY}, the GHY term and its relation to horizon entropy will be introduced. Next section will deal with the construction for the main formalism based on purely surface term. Here the entropy function will be constructed from the GHY action. The whole formalism will be applied in section \ref{Entropy} to find the entropy of the extremal near horizon Ressiner-Nordstrom solution. Final section will summarize the results and then conclude.

\section{\label{GHY}GHY surface term and relation to entropy}
     The usual EH action contains both first order and second order derivatives of the metric $g_{ab}$. As a result, the arbitrary variation of the action leads to a boundary term which is composed of variations of the metric as well as the derivative of the metric. Therefore, to obtain the equation of motion by least action principle one has to impose both the Dirichlet and Neumann boundary conditions; i.e. we have to fix metric and derivative of the metric simultaneously at the boundary. Such a prescription is not well posed in physics. To avoid this discrepancy, people adds a surface term which helps to get rid of this kind of issue. Of course, there is no unique choice of the boundary term \cite{Charap:1982kn}. Most of the cases, people choose the GHY surface term, given originally by Gibbons, Hawking and York. This is defined on the timelike or spacelike surfaces and hence it is foliation dependent {\footnote{An attempt has been made recently in \cite{Parattu:2015gga} to define a boundary term on an arbitrary null surface.}}. Inclusion of it leads to the fact that we have to fix only the induced metric, defined on the foliated surface, at the boundary to find the Einstein's equations of motion \cite{Paddy}. In this section, introducing the GHY term, I shall briefly discuss the relations between it and the horizon entropy of a black hole.
     
     The GHY surface term is given by
\begin{equation}
\mathcal{A}_{GHY} = -\frac{1}{8\pi G}\int_{\partial \mathcal{V}}\sqrt{|h_{(i)}|}d^3x \epsilon K_{(i)}~,
\label{3.01}
\end{equation}
where $\partial \mathcal{V}$ is the three dimensional boundary surface of the four dimensional curved manifold $\mathcal{M}$. It can be timelike or spacelike, depending on the value of $\epsilon$. $h_{(i)}$ is the determinant of the induced metric on $\partial\mathcal{V}$ and $K_{(i)}$ is the second fundamental which is expressed in terms of the unit normal $N^a_{(i)}$ to the boundary as 
\begin{equation}
K_{(i)}=-\nabla_aN^a_{(i)}~.
\label{3.02}
\end{equation}
Here ``$i$'' in the subscript denotes the kind of surface (spacelike or timelike) we are choosing and $\epsilon =+1$ for timelike surface while $\epsilon=-1$ for spacelike surface.
Using Gauss's theorem (\ref{3.01}) can be expressed as
\begin{equation}
\mathcal{A}_{GHY} = -\int_{\mathcal{M}}d^4x\sqrt{-g}\mathcal{L}_{GHY} =-\frac{1}{8\pi G}\int_{\mathcal{M}}d^4x\sqrt{-g} \nabla_a\Big(K_{(i)}N^a_{(i)}\Big)~,
\label{3.03}
\end{equation}
where $N^a_{(i)}N_{a(i)}=\epsilon$ has been used.
For the present context ``$i$'' denotes $t=constant$ surface which is spacelike and $r=constant$, $\theta=constant$, $\phi=constant$ surfaces which are timelike.

  Now it is well known that if one calculates the surface term (\ref{3.01}) on a $r=constant$ surface and then takes the horizon limit, it leads to the black hole entropy. To illuminate this fact, let me give a brief discussion on this. Consider, for simplicity, a static spherically symmetric non-extremal black hole of the form:
\begin{equation}
ds^2 = -f(r)dt^2+\frac{dr^2}{l(r)}+r^2d\Omega^2~,
\label{3.04}
\end{equation}
whose horizon is given by $f(r_H)=0=l(r_H)$. For the $r=constant$ surface, integration in (\ref{3.01}) will be on $t,\theta$ and $\phi$. Here the determinant of the induced metric is $h_{(r)}=-r^4f(r)\sin^2\theta$ and the trace of the second fundamental turns out to be $K_{(r)}= -(f'\sqrt{l})/(2f)-(2/r)\sqrt{l}$ which is independent of the coordinates of the surface. Also note that the periodicity of the Euclidean time is given by $(2\pi)/\kappa$ where the surface gravity $\kappa = \sqrt{f'(r_H)l'(r_H)}/2$. Using the above information and taking $\epsilon=+1$ (as the $r=constant$ surface is timelike) we obtain the following value of the GHY term in the near horizon limit:
\begin{equation}
\mathcal{A}_{GHY}|_{\mathcal{H}} = -\frac{1}{8\pi G}\Big[K_{(r)}\sqrt{f}\Big]_{\mathcal{H}}\frac{2\pi}{\kappa}A_H = \frac{A_H}{4G}~. 
\label{3.05}
\end{equation}
The integration range of time is taken to be the periodicity of the Euclidean time and $A_H$ is the horizon area. This shows that the GHY surface term is related to the horizon entropy. The connection between the GHY term and entropy of horizon is not new. It has been observed earlier that the contribution to the partition function, which leads to the entropy of a black hole, in the Euclidean approach, in the case of Schwarzschild metric comes soley from the surface term as the $R$ vanishes \cite{Gibbons:1976ue}. Such connection has also been discussed in \cite{Iyer:1995kg}. On the other hand, $\mathcal{A}_{GHY}|_{\mathcal{H}}$ can also be interpreted as the surface Hamiltonian by using the Hamilton-Jacobi result $H_{sur}=-(\partial \mathcal{A}_{GHY}|_{\mathcal{H}}/\partial t)$ \cite{Majhi:2013jpk}. This turns out to be $H_{sur}=TS_{BH}$ where $T=\kappa/2\pi$ is the Hawking temperature and $S_{BH}=A_H/4G$ is the black hole entropy. 

   Also it has been shown that the Noether charge corresponding to GHY term, calculated on the horizon for the timelike Killing vector, leads to entropy when multiplied by the periodicity of the Euclidean time \cite{Majhi:2012tf,Padmanabhan:2012bs}. The anti-symmetric potential for any diffeomorphism $x^a\rightarrow x^a+\xi^a$ is given by
\begin{equation}
J^{ab}[\xi]=\frac{1}{8\pi G}K_{(i)}(\xi^aN^b_{(i)}-\xi^bN^a_{(i)})~.
\label{2.03}
\end{equation}
Now calculation of the charge on the horizon for the metric (\ref{3.04}) leads to
\begin{equation}
Q_H = \frac{1}{2}\int_{\mathcal{H}}\sqrt{\sigma}d\Sigma_{ab}J^{ab}[\chi]=\frac{\kappa A_H}{8\pi G}
\label{2.04}
\end{equation}
where $\chi^a=(1,0,0,0)$ is the timelike Killing vector and $\sigma$ is the determinant of the induced metric on the horizon. Multiplying it by the periodicity of the Euclidean time one obtains the entropy $S_{BH}=(2\pi/\kappa)Q_H = A_H/4G$.

   The discussion, presented in this section, tells that the GHY surface term plays a major role to study the thermodynamics of gravity. In fact, many physical entities and several information of the theory can be extracted from this without any information of the bulk term. This has been already mentioned in the introductory part. In the next section I shall discuss the role of the surface term in the context of the entropy function and entropy of a ENH-BH.

\section{\label{Extremal}Extremal near horizon black hole and entropy function from GHY term}
  To start with, let me introduce a brief discussion on the ENH-BH. Consider a theory where a four dimensional gravity theory is coupled to $U(1)$ gauge fields $A_a^{(j)}$ and neutral scalar fields $\{\phi_s\}$. So the total Lagrangian for our theory is $\mathcal{L}=\mathcal{L}_{g}+\mathcal{L}_m$ where first one is the gravity part while the last one is the matter part. In this case $\mathcal{L}_m$ is given by
\begin{equation}
\mathcal{L}_m = -\frac{1}{4}F_{ab}^{(j)}F^{(j)ab}+\textrm{Lagrangian for the scalar fields}~,
\label{2.01}
\end{equation}
with $F^{(j)}_{ab}=\nabla_a A_{b}^{(j)}-\nabla_b A_{a}^{(j)}$.
One spacial solution of this kind of theory is Reissner-Nordstrom (RN) black hole metric.

  The extremal near horizon black hole solution, in general, takes the form as $AdS_2\times S^2$. It is invariant under $SO(2,1)\times SO(3)$ transformations. In four spacetime dimensions, one writes in the following way:
\begin{eqnarray}
&&ds^2 = v_1\Big(-r^2 dt^2+\frac{dr^2}{r^2}\Big) + v_2 (d\theta^2+\sin^2\theta d\phi^2)~;
\nonumber
\\
&&\phi_s=u_s;
\nonumber
\\
&&F^{(j)}_{rt}=e_j; \,\,\,\ F^{(j)}_{\theta\phi} = \frac{p_j}{4\pi}\sin\theta~,
\label{2.02}
\end{eqnarray}
where $v_1,v_2,\{u_s\},\{e_j\}$ and $\{p_j\}$ are constants. For $U(1)$ case, $e_j$ and $p_j$ are electric field and magnetic charge, respectively. A detailed analysis to obtain the ENH-BH for RN solution is presented in \cite{Sen:2007qy}.

   Now define a function $F(\vec{u},\vec{v},\vec{e},\vec{p})$ by integrating the total Lagrangian density on the transverse (angular) coordinates for the extremal near horizon geometry (\ref{2.02}). It has been mentioned in the introduction that one can obtain the Einstein's equations of motion from the total action which is composed of the GHY surface term and the matter part. Therefore (\ref{2.02}) can be considered as a solution of the theory which is given by the total action $\mathcal{L} = \mathcal{L}_{GHY}+\mathcal{L}_m$ with $\mathcal{L}_{GHY}= 1/(8\pi G)\nabla_a\Big(K_{(i)}N^a_{(i)}\Big)$ and $\mathcal{L}_m$ is the matter action, given by (\ref{2.01}). Keeping this in mind we define the function as 
\begin{equation}
F(\vec{u},\vec{v},\vec{e},\vec{p})=\int d\theta d\phi \sqrt{-g}(\mathcal{L}_{GHY}+\mathcal{L}_{m})~.
\label{3.06}
\end{equation}
Use of the gauge field equations and the Bianchi identities for the full black hole solution lead to the following important results \cite{Sen:2005wa,Sen:2007qy}:
\begin{equation}
\frac{\partial F}{\partial e_j}=q_j~,
\label{3.15}
\end{equation}
where the constant $q_j$ is identified as the electric charge and $p_j$ is the magnetic charge of the black hole. Also since the Einstein's equations of motion can be obtained by extremizing the GHY plus the matter action, it is obvious that the scalar and the metric field equations for the extremum near horizon geometry correspond to the extremization of $F$ with respect to $\vec{u}$ and $\vec{v}$, respectively; i.e.
\begin{equation}
\frac{\partial F}{\partial u_s}=0; \,\,\,\ \frac{\partial F}{\partial v_j}=0~.
\label{3.16}
\end{equation}

 In order to find a relation between $F$ and the entropy of the black hole, we define $F_\lambda(\vec{u},\vec{v},\vec{e},\vec{p})$ by rescaling $K_{(r)}$ by $\lambda$; i.e. replacing $K_{(r)}$ by $\lambda K_{(r)}$ with $\lambda$ is a constant. Then taking differentiation on both sides with respect to $\lambda$ and finally putting $\lambda=1$, we obtain
\begin{eqnarray}
\frac{\partial F_\lambda(\vec{u},\vec{v},\vec{e},\vec{p})}{\partial\lambda}\Big{|}_{\lambda=1} = \frac{1}{8\pi G}\int d\theta d\phi \sqrt{-g} \nabla_a\Big(K_{(r)}N^a_{(r)}\Big)~.
\label{3.07}
\end{eqnarray}  
Since for the metric (\ref{2.02}), only non-vanishing component of $N^a_{(r)}$ is radial component, the above reduces to
\begin{equation}
\frac{\partial F_\lambda(\vec{u},\vec{v},\vec{e},\vec{p})}{\partial\lambda}\Big{|}_{\lambda=1} = \frac{1}{8\pi G}\int d\theta d\phi \partial_r\Big(K_{(r)}N^r_{(r)}\sqrt{-g}\Big)~.
\label{3.08}
\end{equation}
Now for this metric one finds $N^r_{(r)}=r/\sqrt{v_1}$ and $K_{(r)}=-1/\sqrt{v_1}$. Substitution of them leads to
\begin{equation}
\frac{\partial F_\lambda(\vec{u},\vec{v},\vec{e},\vec{p})}{\partial\lambda}\Big{|}_{\lambda=1} = \frac{1}{8\pi G} \frac{K_{(r)}}{\sqrt{v_1}}\sqrt{-g_{tt}g_{rr}}\int \sqrt{\sigma}d\theta d\phi = -\frac{A_H}{8\pi G}~.
\label{3.09}
\end{equation}
Here $\sigma$ is the determinant of the induced metric on $t$ and $r$ constant surface and the integration is performed on the horizon.
Hence we can express the black hole entropy as
\begin{equation}
S_{BH} = \frac{A_H}{4G}=-2\pi \frac{\partial F_\lambda(\vec{u},\vec{v},\vec{e},\vec{p})}{\partial\lambda}\Big{|}_{\lambda=1}~.
\label{3.10}
\end{equation}

      Next consider following function by taking the Legendre's transformation of $F(\vec{u},\vec{v},\vec{e},\vec{p})$:
\begin{equation}
E(\vec{u},\vec{v},\vec{q},\vec{p})=2\pi\Big[e_jq_j - F(\vec{u},\vec{v},\vec{e},\vec{p})\Big]~. 
\label{3.11}
\end{equation} 
In terms of $E$ the equations of motion (\ref{3.16}) and (\ref{3.15}) are given by
\begin{equation}
\frac{\partial E}{\partial u_s}=0; \,\,\ \frac{\partial E}{\partial v_j}=0; \,\,\,\ \frac{\partial E}{\partial e_j}=0~,
\label{3.17}
\end{equation}
respectively.
In the following using (\ref{3.10}) and (\ref{3.17}) I shall relate $E$ with the black hole entropy.
To do this, note that the Lagrangian $\mathcal{L}_\lambda$, which is achieved by replacing $K_{(r)}$ by $\lambda K_{(r)}$ in the total Lagrangian (see Eq. (\ref{3.06})), is invariant under the scaling $\lambda\rightarrow s\lambda$, $v_1\rightarrow sv_1$ and $e_j\rightarrow se_j$ where $s$ is an arbitrary constant. Therefore, since $\sqrt{-g}\sim v_1$, the function $F_\lambda(\vec{u},\vec{v},\vec{e},\vec{p})$ is scaled as $sF_\lambda(\vec{u},\vec{v},\vec{e},\vec{p})$. Hence $F_{\lambda}$ is a function of $\lambda,v_1$ and $e_j$ with degree one. So using Euler's theorem, we can write:
\begin{equation}
F_\lambda(\vec{u},\vec{v},\vec{e},\vec{p})=\lambda\frac{\partial F_\lambda(\vec{u},\vec{v},\vec{e},\vec{p})}{\partial\lambda}+v_1\frac{\partial F_\lambda(\vec{u},\vec{v},\vec{e},\vec{p})}{\partial v_1}+e_j\frac{\partial F_\lambda(\vec{u},\vec{v},\vec{e},\vec{p})}{\partial e_j}~.
\label{3.12}
\end{equation}
Then taking $\lambda=1$ and using (\ref{3.15}) and the equations of motion (\ref{3.16}), one obtains
\begin{equation}
\frac{\partial F_\lambda(\vec{u},\vec{v},\vec{e},\vec{p})}{\partial\lambda}\Big|_{\lambda} = F(\vec{u},\vec{v},\vec{e},\vec{p}) - e_jq_j~.
\label{3.13}
\end{equation} 
Hence use of (\ref{3.10}) and (\ref{3.11}) yields
\begin{equation}
S_{BH}=2\pi\Big[e_jq_j - F(\vec{u},\vec{v},\vec{e},\vec{p})\Big] = E(\vec{u},\vec{v},\vec{q},\vec{p})~.
\label{3.14}
\end{equation}
This implies the entropy of the extremal near horizon black hole is given by the value of $E(\vec{u},\vec{v},\vec{q},\vec{p})$ at the extremum with the extremization is done by the set of equations (\ref{3.17}).

    So the calculation of entropy for an extremal near horizon black hole reduces to a set of algebraic equations which are given by the relations, presented in (\ref{3.17}). The steps are as follows. First calculate $F(\vec{u},\vec{v},\vec{e},\vec{p})$ using (\ref{3.06}) to obtain $E(\vec{u},\vec{v},\vec{q},\vec{p})$. Then extremize it, basically leads to (\ref{3.17}), which in turn yields a set of algebraic equations. Use them back into the expression for $E$ which yields the value of $E$ at the extremum. This gives the entropy of the black hole. In the following I shall use this setup to find this for the metric (\ref{2.02}).

\section{\label{Entropy}Entropy from the entropy function}
 To calculate $F$ we need to evaluate $\mathcal{L}_{GHY}$ and $L_m$. Here consider that (\ref{2.02}) is the solution of the theory where the matter part is given by the action for the gauge fields only; i.e. it represents an extremal near horizon RN black hole. First concentrate on the GHY part. To proceed further write it in the following form:
\begin{eqnarray}
\sqrt{-g}\mathcal{L}_{GHY} &=& \frac{1}{8\pi G}\partial_a\Big(\sqrt{-g}K_{(i)}N^a_{(i)}\Big)
\nonumber
\\
&=& \frac{1}{8\pi G}\Big[\partial_a\Big(\sqrt{-g}K_{(t)}N^a_{(t)}\Big)+\partial_a\Big(\sqrt{-g}K_{(r)}N^a_{(r)}\Big)
\nonumber
\\
&+& \partial_a\Big(\sqrt{-g}K_{(\theta)}N^a_{(\theta)}\Big)+\partial_a\Big(\sqrt{-g}K_{(\phi)}N^a_{(\phi)}\Big)\Big]~.
\label{4.01}
\end{eqnarray}
Now, note that the non-vanishing component of the unit normal $N^a_{(t)}$ for the metric (\ref{2.02}) is the temporal part. Similar happens for the other terms; i.e. only radial component of $N^a_{(r)}$ is non-zero and so on. Hence the partial derivative in the first term of the above will be with respect to time. Since the metric is static, this leads to zero. In the identical way, the last term also vanishes. Only the second and the third terms will contribute to the gravity part. Use of $\sqrt{-g}=v_1v_2\sin\theta$, $N^r_{(r)}=r/\sqrt{v_1}$, $K_{(r)}=-1/\sqrt{v_1}$, $N^\theta_{(\theta)} = 1/\sqrt{v_2}$ and $K_{(\theta)} = -\cos\theta/(\sqrt{v_2}~\sin\theta)$ yields,
\begin{equation}
\sqrt{-g}\mathcal{L}_{GHY} = \frac{1}{8\pi G}\Big(-v_2\sin\theta + v_1\sin\theta\Big)~.
\label{4.02}
\end{equation}
On the other hand, the matter part leads to
\begin{equation}
\sqrt{-g}\mathcal{L}_m = v_1v_2\sin\theta\Big[\frac{e^2}{2v_1^2}-\frac{1}{2v^2_2}\Big(\frac{p}{4\pi}\Big)^2\Big]~.
\label{4.03}
\end{equation}
Substituting them in (\ref{3.06}) and integrating on the angular variables we obtain
\begin{equation}
F(\vec{u},\vec{v},\vec{e},\vec{p})=\frac{1}{2G}(v_1-v_2)+2\pi\Big[\frac{e^2v_2}{v_1}-\frac{v_1}{v_2}\Big(\frac{p}{4\pi}\Big)^2\Big]~.
\label{4.04}
\end{equation}
Therefore, by (\ref{3.14}), $E$ turns out to be
\begin{equation}
E(\vec{u},\vec{v},\vec{q},\vec{p}) = 2\pi\Big[qe - \frac{1}{2G}(v_1-v_2)-2\pi\Big\{\frac{e^2v_2}{v_1}-\frac{v_1}{v_2}\Big(\frac{p}{4\pi}\Big)^2\Big\}\Big]~.
\label{4.05}
\end{equation}
Then (\ref{3.17}) leads to the following set of equations:
\begin{eqnarray}
&&-\frac{1}{2G}+\frac{2\pi e^2v_2}{v_1^2}+\frac{p^2}{8\pi v_2}=0~;
\nonumber
\\
&&\frac{1}{2G}-\frac{2\pi e^2}{v_1}-\frac{v_1p^2}{8\pi v_2^2}=0~;
\nonumber
\\
&&q-\frac{4\pi ev_2}{v_1}=0~.
\label{4.06}
\end{eqnarray}
Solutions of these are
\begin{equation}
e=\frac{q}{4\pi}; \,\,\,\ v_1=v_2=G\frac{q^2+p^2}{4\pi}~.
\label{4.07}
\end{equation}
Substitution of the above values in (\ref{4.05}) leads to the value of $E$ at extremum, which is by (\ref{3.14}) is the entropy of the black hole:
\begin{equation}
S_{BH} = E|_{extremum} = \frac{1}{4}(q^2+p^2)~.
\label{4.08}
\end{equation}
This shows that an entropy function for the ENH-BH solutions can be constructed only from the GHY term which leads to the correct value of the horizon entropy.

\section{\label{Conclusions}Summary and Conclusions}
   In stead of looking at the full action, it has been demonstrated that the whole entropy function formalism can be developed just by considering the GHY surface term. The steps, adopted here, are identical to the original work of Sen \cite{Sen:2005wa,Sen:2007qy}. The only difference occurred here in the action for the theory. I never borrowed any information of the main action, like EH action, in the sense that everything has been constructed based on the pure surface term. Another interesting feature to be noted that in defining the function $F$ (see Eq. (\ref{3.06})), the Lagrangian density for GHY is taken to be covariant derivative over all coordinates. Usually GHY is defined on any timelike or spacelike surface; whereas in Eq. (\ref{3.06}) the second fundamental and the normals are defined for the manifold whose boundary consists of one spacelike and three timelike surfaces. The correct result is coming when one considers all the contribution (see the analysis around Eq. (\ref{4.02})). A similar feature has been observed earlier \cite{Padmanabhan:2002sha} in interpreting the gravitational action as the inverse temperature times the free energy. 

  Earlier several instances showed that the surface term may reflects the most of the information of the bulk. So it would be interesting to investigate if there is any role of the surface term in the context of entropy function formalism. Here I precisely addressed this question and found that it is indeed possible. This again strengthens the fact that in gravitational theory these terms play a major part in the dynamics of the gravity in the sense that one can find the equations of motion by extremizing the entropy. Finally, the formalism is general enough to investigate for a general Lanczos-Lovelock theory. This is in progress \cite{Bibhas}.

\vskip 9mm
\noindent
{\bf{Acknowledgements}}\\
\noindent
The research of the author is supported by a START-UP RESEARCH GRANT (No. SG/PHY/P/BRM/01) from Indian Institute of Technology Guwahati, India.

\end{document}